\documentclass[twocolumn]{aastex631}
\usepackage[T1]{fontenc}
\usepackage{hyperref}
\usepackage{amsmath}
\usepackage{amssymb}
\usepackage{amsfonts}
\usepackage{multirow} 
\usepackage{booktabs}

\newcommand\aastex{AAS\TeX}

\usepackage{hyperref}
	\hypersetup{
    unicode=true,          
    pdftoolbar=true,        
    pdfmenubar=true,        
    pdffitwindow=false,     
    pdfstartview={FitH},    
    pdftitle={My title},    
    pdfauthor={Author},     
    pdfsubject={Subject},   
    pdfcreator={Creator},   
    pdfproducer={Producer}, 
    pdfkeywords={keyword1, key2, key3}, 
    pdfnewwindow=true,      
    colorlinks=true,        
    linkcolor=MidnightBlue,     
    citecolor=MidnightBlue,     
    filecolor=MidnightBlue,     
    urlcolor=MidnightBlue       
}

\newcommand{\zem}{$z_{\rm em}$ }
\newcommand{\zabs}{$z_{\rm abs}$}
\newcommand{\kms}{$\rm km~s^{-1}$}
\newcommand{\hi}{\mbox{\tiny H\,{\sc i}}}
\newcommand{\HI}{\mbox{H\,{\sc i}}}

\def\zhi{$z_{\rm HI}$}
\def\zmgii{$z_{\rm \tiny{MgII}}$}
\def\nhi{$N$($\hi$)}
\newcommand{\mgii}{\mbox{Mg\,{\sc ii}}}

\newcommand{\feii}{\mbox{Fe\,{\sc ii}}}

\newcommand{\caii}{\mbox{Ca\,{\sc ii}}}
\newcommand{\nai}{\mbox{Na\,{\sc i}}}

\newcommand{\MgI}{\mbox{Mg\,{\sc i}}}
\newcommand{\MgII}{\mbox{Mg\,{\sc ii}}}

\newcommand{\FeII}{\mbox{Fe\,{\sc ii}}}

\newcommand{\CaII}{\mbox{Ca\,{\sc ii}}}

\begin{document}
\title{Template \aastex v7 Article with Examples\footnote{Footnotes can be added to titles}}

\title{Multiple H\textsc{i} 21-cm absorbers towards blazar PKS 1158+007}

\author[0000-0001-6047-4521]{Abhisek Mohapatra}
\affiliation{Department of Astronomy, University of Cape Town, Private Bag X3, University Avenue N, Rondebosch, Cape Town, 7700, South Africa}

\author[0000-0002-9062-1921]{Raghunathan Srianand}
\affiliation{IUCAA, Postbag 4, Ganeshkhind, Pune 411007, Maharastra, India}

\author[0000-0002-6095-7627]{Rajeshwari Dutta}
\affiliation{IUCAA, Postbag 4, Ganeshkhind, Pune 411007, Maharastra, India}

\correspondingauthor{Abhisek Mohapatra}
\email{abhisekphy@gmail.com}




\begin{abstract}
We report an interesting line of sight towards the blazar PKS 1158+007 (J1201+0028) (\zem\ = 1.369), revealing the rare detection of two intervening \HI\ 21-cm absorption systems within a narrow redshift interval ($\Delta z \sim 0.04$) using the MeerKAT radio telescope. The stronger absorber at \zabs\ $=0.5653$ ($\int\tau dv = 1.31\pm0.04$ \kms) is associated with a strong \mgii\ $\lambda2796$ system seen in the Sloan Digital Sky Survey (SDSS) and Dark Energy Spectroscopic Instrument (DESI) spectra, with rest-frame equivalent widths of $W_{2796} = 1.40\pm0.07$~\AA\ and $1.21\pm0.11$~\AA, respectively.
 In stark contrast, the second detection at \zabs\ $= 0.5305$ reveals a distinct \HI\ 21-cm absorption feature ($\int\tau dv = 0.31 \pm 0.04$ \kms) with no associated metal absorption detectable in the optical spectra. This system represents a prime example of a ``metal-poor'' \HI\ 21-cm absorber. Using photometric redshifts from Hyper Suprime-Cam (HSC) PDR3 and DESI Legacy Surveys DR10, we identify a broad galaxy overdensity (at $z\
 \sim0.53\pm0.06$) coincident with these absorbers, suggesting the sightline pierces a complex large-scale structure or filament, which would be interesting to confirm with follow-up observations. These results underscore the power of blind \HI\ 21-cm surveys to uncover the ``invisible'' cold gas phase of the Universe that optical/UV metal-line selection may systematically miss.
\end{abstract}



\section{Introduction}
\label{sec:intro}

\HI\ 21-cm absorption is an excellent probe of the physical conditions prevailing in the cold neutral gas. Unlike other tracers of \HI\ gas such as Damped Lyman-$\alpha$ (DLA, i.e., $ {\rm log}[N(\HI)/{\rm cm}^{-2}] \ge 20.3$) systems and \MgII\ systems seen in the optical spectra of distant quasars, intervening \HI\ 21-cm absorption is very rare. For example, roughly $\sim10\%$ of $z\gtrsim2$ DLAs show detectable \HI\ 21-cm absorption \citep[see][]{Srianand12, Kanekar14}. Similarly, the detection rate of \HI\ 21-cm absorption towards strong \MgII\ (rest equivalent width of the \mgii\ $\lambda 2796$, $\rm W_{2796} \ge 1$ \AA) systems is found to be small \citep[i.e., $\sim10-20$\%; see][]{Kanekar09mg2,Gupta09,Gupta12,Dutta17mg2,Dutta17fe2}. The small detection rates are understood to be due to either the neutral gas having a higher spin temperature ($T_S$), or the covering factor of cold gas (i.e., the ratio of the projected cross-section of the cold gas and that of the background radio continuum source) being small. 

Search for \HI\ 21-cm absorption along lines of sight close to $z\lesssim0.4$ galaxies (i.e., quasar-galaxy pairs) also resulted in small detection rates \citep[$\approx24$\% within 15 kpc; see][]{Gupta10, Dutta17}, with the detection rate showing a strong declining trend beyond 10 kpc \citep[][]{Guha2025}. On the other hand, the detection rate of DLAs around galaxies shows a shallower declining trend with impact parameter compared to \HI\ 21-cm absorption \citep{kulkarni2022}. This is consistent with the \HI\ gas around galaxies being patchy, and the cold \HI\ phase suitable for producing 21-cm absorption having a very small cross-section. 

Existence of small scale fluctuation in the \HI\ column density, $N$(\HI), and spin-temperature is well documented in the case of the Galactic Interstellar medium \citep[ISM;][]{Frail94, Heiles97, Brogan05, Roy12}. However, such observations are rare in the extragalactic setting \citep[see however,][]{Wolfe82, Kanekar01, Srianand2013, Dutta16,Gupta_2018, Srianand2022}. Radio sources showing strong proper motions and/or structures over a wide range of projected distance scales are ideal for probing the structures in the intervening absorbing gas.

Here, we report the detection of two \HI\ 21-cm absorptions within a narrow redshift interval of $\Delta z \sim 0.04$ towards a blazar PKS 1158+007 (J1201+0028) that shows core-jet structure with strong variability in all wavebands. As we discuss here, this is a low probability occurrence given the fact
that \HI\ 21-cm absorption detections are very rare. 
One of the detected \HI\ 21-cm absorption is associated with a strong \MgII\ system detected in the optical spectra of the quasar. Interestingly, the other one has no associated metal absorption (from \MgII, \MgI, \FeII\ or \CaII) detected in the optical quasar spectra. 
While metal poor neutral gas has been detected \citep[see e.g.,][]{Welsh2024}, this is the first reported case of a metal-poor \HI\ 21-cm absorption. Here, we present a detailed analysis of this interesting line of sight.

This paper is organized as follows. In section~\ref{sec:blazar}, we provide the details of the target and absorption systems detected in the optical spectra. In section~\ref{sec:obsmrkt}, we provide the details of the MeerKAT observations and data reduction. Details of the \HI\ 21-cm absorbers and their association to the metal line systems are discussed in section~\ref{sec:21cm}.
Discussions and conclusions are presented in section~\ref{sec:summary}. In this paper, we consider a flat Universe with the following cosmological parameter,  $\Omega_m$=0.3, $\Omega_\Lambda$=0.7 and H$_{\rm 0}$=70\,\kms\,Mpc$^{-1}$. All the positions are provided in J2000 equatorial coordinates. 

\section {The Blazar J1201+0028}
\label{sec:blazar}

\begin{table}[ht]
\centering
\scriptsize
\caption{
Rest-frame equivalent widths of Mg\,\textsc{ii} and Fe\,\textsc{ii} absorption lines.
Upper limits provided are 3$\sigma$ limits assuming a Full Width at Half Maximum (FWHM) of
200 km s$^{-1}$.
}
\label{tab:mgii_feii_comparison}
\renewcommand{\arraystretch}{1.25}
\begin{tabular}{l l cc}
\toprule
 & \multicolumn{2}{c}{Mg\,\textsc{ii}} & Fe\,\textsc{ii} \\
\cmidrule(lr){2-3} \cmidrule(lr){4-4}
 Spectrum$^*$ 
& $W_r(2796)$ (\AA) 
& $W_r(2803)$ (\AA) 
& $W_r(2600)$ (\AA) \\
\midrule
\multicolumn{4}{c}{\zabs = $0.50046$ system}\\
SDSS-1 & $2.02 \pm 0.05$ & $0.84 \pm 0.04$ & $0.74 \pm 0.01$ \\
SDSS-2 & $1.28 \pm 0.06$ & $1.01 \pm 0.06$ & $0.21 \pm 0.01$ \\
DESI   & $1.50 \pm 0.08$ & $1.07 \pm 0.07$ & $0.64 \pm 0.10$ \\
\midrule
\multicolumn{4}{c}{\zabs = 0.5305 system}\\
SDSS-1 & $<0.41$ & $<0.40$ & $<0.40$ \\
SDSS-2 & $<0.19$ & $<0.18$ & $<0.18$ \\
DESI   & $<0.12$ & $<0.12$ & $<0.10$ \\
\midrule
\multicolumn{4}{c}{\zabs = 0.56544 system}\\
SDSS-1 & $1.40 \pm 0.07$ & $1.16 \pm 0.08$ & $0.68 \pm 0.03$ \\
SDSS-2 & $1.12 \pm 0.08$ & $0.87 \pm 0.08$ & $0.40 \pm 0.03$ \\
DESI   & $1.21 \pm 0.11$ & $1.08 \pm 0.11$ & $0.66 \pm 0.10$ \\
\bottomrule
\end{tabular}
\medskip
\begin{minipage}{0.85\linewidth} 
\footnotesize
\textbf{Note --} *: Observation Modified Julian Dates (MJD) for the epochs are: SDSS-1 = 51663; SDSS-2 = 51930; DESI = 59273. Typical $1\sigma$ error in the redshift is 0.00006.
\end{minipage}

\end{table}

The source J1201+0028 (also referred to as PKS 1158+007) at \zem = 1.36874 is classified as a blazar based on its multi-band colors \citep[see for example,][]{d'abrusco2014,Massardi2022}. The strong emission lines present in the 
Sloan Digital Sky Survey (SDSS) and Dark Energy Spectroscopic Instrument (DESI) spectra confirm this as a Flat Spectrum Radio Quasar (FSRQ) class of Blazars. 
The quasar continuum spectrum shows strong variability between the three epochs with a tendency to become redder when brighter. The Zwicky Transient Facility (ZTF) light curves of J1201+0028 in r and g bands show strong ($\sim$0.3 mag) variability over short time-scales ($\sim 85$ days in the quasar's rest frame) as well.

This source is known to have a flat spectrum (spectral index, $\alpha = 0.26$) radio core with one-sided extended emission at arcsecond scales \citep[see][]{Downes1986}. The L-band (1.4 GHz) observations from the Faint Images of the Radio Sky at Twenty-Centimeters (FIRST) survey \citep{Becker95} suggest three components spread over $\sim$10 arcsec with the strongest component having $\sim95$\% of the total flux density. The Gaussian fit to the strongest component also suggests an elongated feature with the measured peak flux density contributing to $\sim$89\% of the total flux density of this component. { Strong L-band flux variability is also noticed between observations, as given in the NASA/IPAC EXTRAGALACTIC DATABASE (NED)}. In the L-band radio maps obtained for the FIRST survey, the location of the peak flux density shows an offset of about 1 arcsec (which is higher than the typical positional uncertainty of $\le 0.1$ arcsec for sources brighter than 10 mJy) with respect to the optical positions from SDSS and other ground-based surveys.  At $z\sim0.5$, this will correspond to a projected separation of $\sim$6 kpc.

The Very Long Baseline Array (VLBA) images  (at 2.3 and 9.6 GHz) presented by \citet{Kovalev2007} show the source to be compact at milliarcsecond (mas) scales with $\sim$92\% of the total flux density associated with the peak emission. We also note from the available VLBA X-band (8.7 GHz) and S-band (2.3 GHz) maps at the Astrogeo VLBI FITS image database\footnote{https://astrogeo.org/vlbi\_images/} that J1201+0028 shows a core-jet morphology \citep[with a separation of $4.5\pm0.9$ mas;][]{Plavin2022}, with the total flux measured at the mas scale showing variability. 
\citet{Petrov2017} found an offset of 2.72 mas between the VLBA and Gaia positions. However, this was not found to be statistically significant considering the allowed uncertainties in the VLBA and Gaia position measurements \citep[see also][]{Valeri2019}.

The total flux density measured in the VLBA images is less than 50\% of what is expected based on the available measurements at arcsec scales. This implies that a significant fraction of the radio emission seen at arcsec scale is resolved out in the VLBA images. 
Thus, the source has structures at all scales (i.e., mas to arcsec scales).  Therefore, the offset found in the L-band images could indicate that the low frequency fluxes (where \HI\ 21-cm absorption is searched) are dominated by the diffuse emission that has an offset with the core seen in the high frequency VLBA images (and optical emission).

\citet{Bouche2006M} have reported two \MgII\ absorbers in the SDSS spectrum of this quasar at \zabs = 0.50046 and 0.56498. The rest-frame equivalent width measurements for \MgII\ and \FeII\ lines are summarised in Table~\ref{tab:mgii_feii_comparison}. These show tentative evidence for variation (typically at $<3\sigma$ level) in the rest-frame equivalent width of absorption lines.

All the above makes J1201+0028 an interesting target to study absorption towards the radio and optical emission.

\section{Details of MeerKAT observations and Data reduction}
 \label{sec:obsmrkt}

The target field, centered on PKS~1158+007 (R.A. $12^{\mathrm{h}}01^{\mathrm{m}}23^{\mathrm{s}}.25$, Decl. $+00^{\circ}28^{\prime}28^{\prime\prime}.3$), was observed using the MeerKAT-64 array \citep{Jonas2016} on 05 Spetember, 2020 (Project ID: $\rm SCI-20180516-NG-01$). The observations utilized the L-band receivers (856--1712 MHz) and the SKA Reconfigurable Application Board (SKARAB) correlator configured in 32K mode, providing a native channel width of 26.123~kHz. The total duration of the observing run (Block ID 1599303664) was 3.49 hours, out of which PKS~1158+007 has an on-source integration time of 55.44 minutes.
%
J1939$-$6342 was observed for primary flux density scaling, delay, and bandpass calibration, while the compact source J1150$-$0023 served as the complex gain (phase) calibrator.


The data reduction was performed using the {\it ilifu} cloud computing facility\footnote{\url{https://www.ilifu.ac.za/}} in Cape Town, utilizing \textsc{Singularity} containers for software management. The raw visibilities were retrieved from the South African Radio Astronomy Observatory (SARAO) archive\footnote{\url{https://archive.sarao.ac.za}} using the \textsc{katdal} library\footnote{\url{https://github.com/ska-sa/katdal}} and converted into Measurement Set (MS) format. 

To prepare the data for spectral line analysis, we applied Doppler correction to the barycentric reference frame using the \textsc{CASA} task \texttt{mstransform} with nearest-neighbor spectral interpolation. While the initial calibration was performed on the full bandwidth, we restricted our subsequent analysis to the frequency range 880--935~MHz to focus on the blind search for redshifted \HI\ 21-cm line of the $z\sim0.5$ \MgII\ absorbers. 

The reduction workflow combined the {\tt{CARACal}} pipeline \citep{Jozsa2020} with custom scripts executed within the {\tt{Stimela2}} framework \citep{Smirnov2025}. The processing strategy consisted of the following main phases. We utilized {\tt{CARACal}} to flag Radio Frequency Interference (RFI) on the calibrators, derive cross-calibration solutions (delay, bandpass, and gain), and apply these to the target. The target data were then flagged for RFI using {\tt{AoFlagger}}\citep{Offringa2012}. 
%
We utilized the \texttt{Stimela2} environment to perform iterative self-calibration. We employed \texttt{WSClean} \citep{Offringa2014} to image the continuum and refine the decontamination masks over several cycles, generating a robust sky model. This model was then used by \texttt{QuartiCal} \citep{Kenyon2025} to derive phase and amplitude gain solutions. Upon convergence, the final self-calibration tables were applied to the MS, and the sky model generated by \texttt{WSClean} was subtracted from the visibilities to isolate the spectral line signal. 
The continuum-subtracted data were imaged to create the final \textsc{H\,i} 21-cm data cube. Any residual continuum emission was removed in the image plane using a spline fit to the visibility spectra by masking the line regions. Finally, the cubes were corrected for the primary beam attenuation.

A summary of the imaging parameters and final noise properties for both the Multi-Frequency Synthesis (MFS) continuum image and the \textsc{H\,i} 21-cm spectral line cube is presented in Table~\ref{tab:obsparam}.

\begin{deluxetable}{lc}

\tablecaption{MeerKAT Imaging Parameters \label{tab:obsparam}}
\tablehead{
    \colhead{Parameter} & 
    \colhead{Value}
} 
\startdata
\multicolumn{2}{c}{\it Continuum Imaging} \\
\hline
Frequency range & $880 - 935$~MHz \\
Imaged area & $8192 \times 8192$ pixels \\
Pixel size & $1\farcs5$ \\
Weighting & Robust = $-0.5$ \\
Synthesized beam & $9\farcs6 \times 9\farcs6$ \\
RMS noise & $34~\mu$Jy beam$^{-1}$ \\
Peak flux density at 0.91~GHz & $521$~mJy beam$^{-1}$ \\
Peak flux density at 0.93~GHz & $511$~mJy beam$^{-1}$\\
\hline
\multicolumn{2}{c}{\it \textsc{H\,i} Spectral Line Imaging} \\
\hline
Frequency range & $880 - 935$~MHz \\
Imaged area & $512 \times 512$ pixels \\
Pixel size & $2\farcs0$ \\
Weighting & Robust = $0.0$ \\
Median Synthesized beam & $14\farcs9 \times 11\farcs6$ \\
RMS noise per channel  & $0.77$~mJy beam$^{-1}$ \\
Channel width & 26.123 kHz \\
\enddata
\tablecomments{The synthesized beam sizes represent the restoring beam used in the final images. }
\end{deluxetable}

\section{Observed properties of \HI\ 21-cm absorbers}
\label{sec:21cm}

\begin{figure*}
\begin{center}
\includegraphics[width=3.5in, trim=0cm 0cm 0cm 0cm, clip]{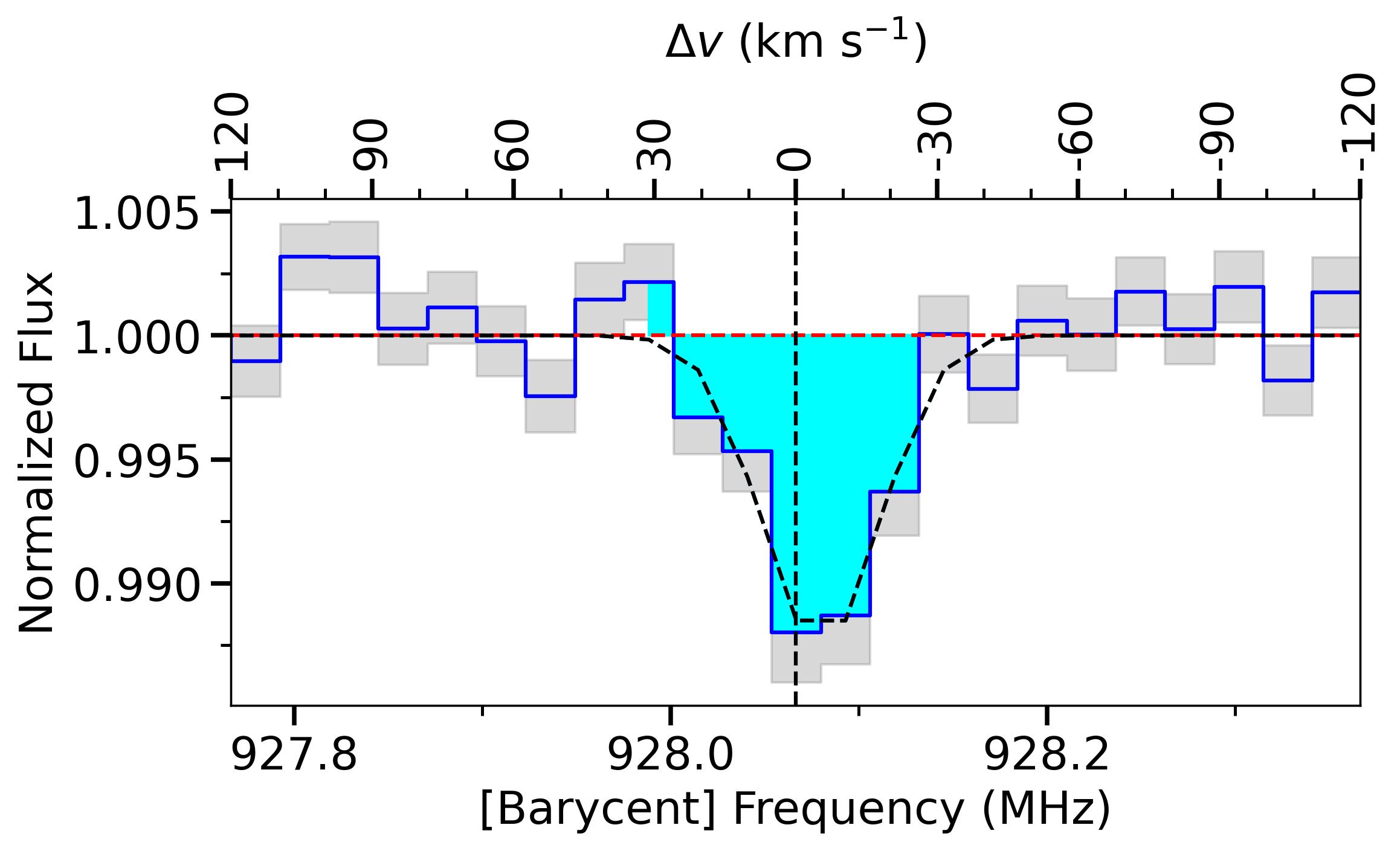}
 \includegraphics[width=3.5in, trim=0cm 0cm 0cm 0cm, clip]{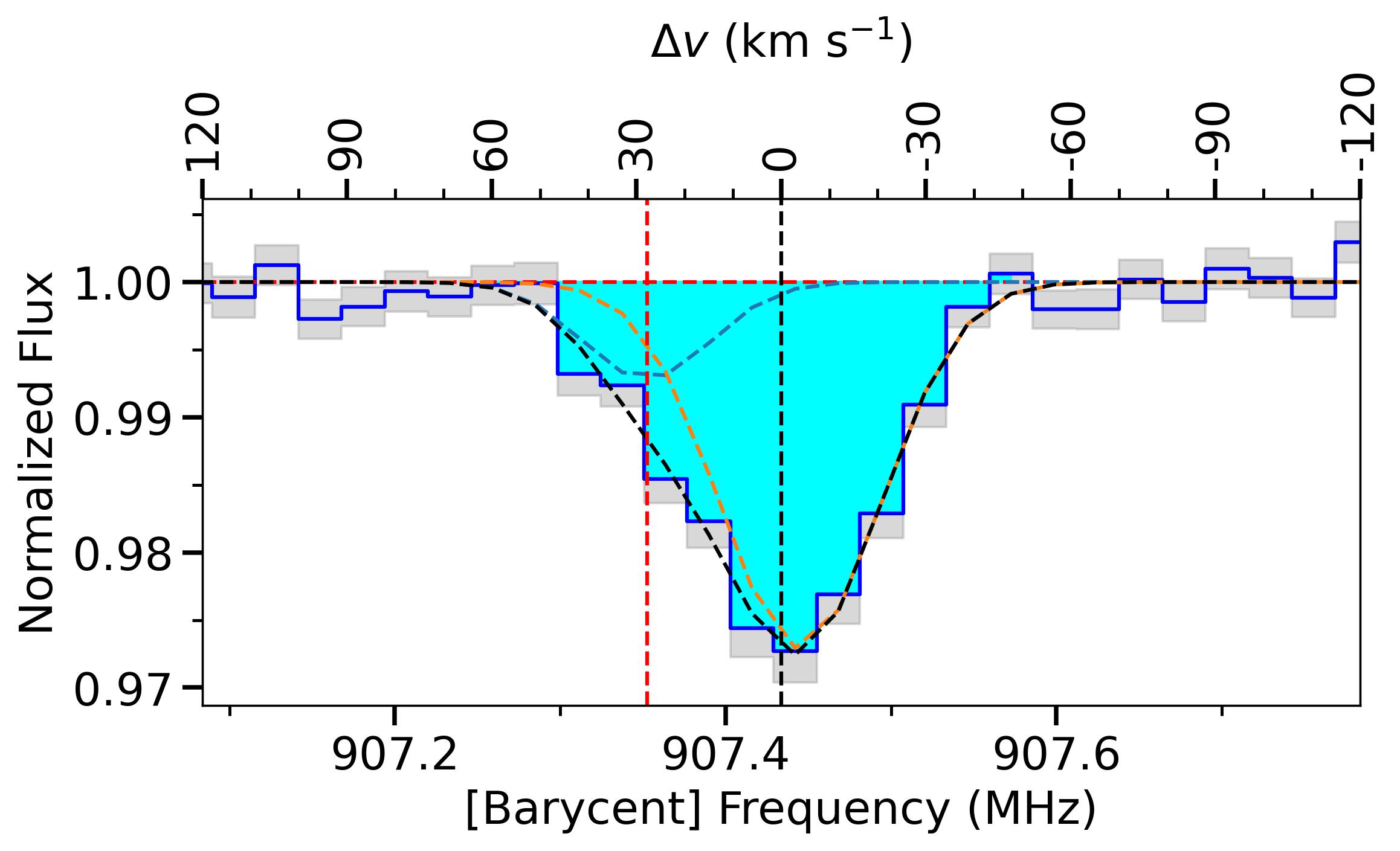}
\caption{\HI\ 21-cm absorptions towards J1201+0028 detected at \zhi=$~ 0.5305$ (left) and at \zhi=$~ 0.5653$ (right). The
spectra are shown at the native spectral resolution of the final
data cube. The gray shaded regions indicate the $1\sigma$ uncertainty in the normalized flux density. The best-fitting Gaussian components are shown as black dashed curves, while color dashed lines indicate the individual Gaussian components. The upper x-axis displays velocities relative to the systemic redshift corresponding to the peak \HI\ 21-cm optical depth. The vertical black dashed line marks zero velocity with respect to the \HI\ 21-cm absorption redshift. In the right panel, the vertical red dashed line indicates the velocity corresponding to the \mgii\ absorption system at $z=0.56544$ relative to \HI\ 21-cm absorption.}
\label{fig:detections}
\end{center}
\end{figure*}
\begin{table}[]
\scriptsize
     \caption{Details of \HI\ 21-cm absorption detected along J1201+0028}
   \centering
    \begin{tabular}{ccccc}
    \hline
    \zabs\,(\zmgii) & \zhi & $\int\tau_{21}$dv & log [$N$(\HI)]~cm$^{-2}$ & $\Delta v_{90}$\\
         &  & (\kms) & &(\kms) \\ 
    \hline
    \hline
   0.50046 & ..... & RFI & .....& .....\\
   \hline
    ..... & 0.53050     & 0.31$\pm$0.04 &  19.75$\pm$0.06 &30.2 \\
   \hline       
    $0.56449^*$ & ..... & $\le 0.06$ & $\le 19.04$& .....\\
    $0.56544^*$ & 0.56530     & 1.31$\pm$0.04 &20.37$\pm$0.01   &62.6 \\
    \hline
    \end{tabular}
    \label{tab:21cm}
    \\
\medskip
\begin{minipage}{0.85\linewidth} 
\footnotesize
\textbf{Note --} $*:$ Two components of \zabs\ = 0.56544 \mgii\ system. \\
We assume $T_S = 100 K$ and $f_c = 1$ to calculate \nhi. The limits are $3\sigma$ and the assumed $dv$ is 10 \kms.
\end{minipage}

\end{table}

We detect two \HI\ 21-cm absorption systems at \zhi\ $= 0.5305$ and $0.5653$ in our MeerKAT spectra (see Figure~\ref{fig:detections}).
The expected frequency range of \HI\ 21-cm absorption corresponding to the \MgII\ absorber at \zabs $ = 0.50046$ is contaminated by strong RFI, preventing a reliable search for the absorption. The \HI\ 21-cm absorber at \zhi $= 0.5653$ coincides with one of the two \MgII\ absorption components of the \MgII\ system at \zabs $ = 0.56544$ seen in the optical spectra (shown in Figure~\ref{fig:optical} of Appendix~\ref{sec:apendix}). In contrast, the \HI\ 21-cm absorber at \zhi $ = 0.5305$ does not show any detectable metal absorption in the available optical spectra. The $3\sigma$ upper limits on the rest equivalent widths of the corresponding \MgII\ and \FeII\ transitions are provided in Table~\ref{tab:mgii_feii_comparison}.

We performed additional tests to assess the robustness of the two \HI\ 21-cm absorption detections against residual RFI. We inspected the frequency-dependent flagging fraction across the observed spectral range. The frequency regions containing the two detected absorbers do not coincide with any strongly flagged regions associated with severe RFI. The \MgII\ system at \zabs = 0.50046, however, lies within a severely RFI-contaminated frequency region and is therefore excluded from the \HI\ 21-cm analysis. The absorption
features also remain present when the data are examined separately in the two parallel-hand polarization products (XX and YY). We further performed a spatial extraction test at random locations away from the quasar to verify that the absorption features do not coincide spatially elsewhere and hence are not due to any imaging artifacts. The details are presented in Appendix~\ref{apexb}.

Assuming that the detected \HI\ 21-cm absorbers completely cover the background radio source and adopting a spin temperature of $T_{\rm S}=100$ K, we estimate the \HI\ column density $N(\HI)$ from the observed absorption profiles. The derived values are summarized in Table~\ref{tab:21cm} (see Column~4). However, as discussed in Section~\ref{sec:blazar}, the covering factor of the absorber is likely to be significantly less than unity, and intervening absorbers are known to have spin temperatures larger than the assumed value of $100$ K \citep[see][]{Kanekar14, Srianand12}. Therefore, the column densities derived here should be considered as lower limits.

The velocity structure of the \HI\ 21-cm absorption provides additional insight into the physical conditions of the neutral gas associated with these systems. We quantify the kinematic extent of the \HI\ 21-cm absorption using $\Delta v_{90}$, measured as the velocity interval between the 5th and 95th percentiles of the cumulative integrated optical-depth profile. The resulting $\Delta v_{90}$ measurements are reported in column (5) of Table~\ref{tab:21cm}. The fitted  Gaussian profile to each detection is shown with a black dashed curve in Figure~\ref{fig:detections}, while the colored dashed curves represent the individual Gaussian components.

\section{Discussions and conclusions}
\label{sec:summary}

The most interesting aspects of these observations are (i) the detection of two \HI\ 21-cm absorbers within a short redshift interval of $\Delta z\sim 0.035$; and (ii) a lack of Mg~{\sc ii} and Fe~{\sc ii} absorption associated with one of these \HI\ 21-cm absorbers. Below we discuss the implications of these observations.

\subsection{On the lack of metal absorption from \zhi = 0.5305 system}

We discuss here the association of \MgII\ absorption with neutral gas probed via DLA systems. \MgII\ absorption is always detected in high-$z$ DLAs whenever the wavelength range is covered \citep{Wolfe2005}. The $z>2$ DLA sample of \citet{Noterdaeme2012} constructed from SDSS has 803 systems with \MgII\ equivalent width measurements, with the lowest W$_r$(2796) value measured being 0.28\,\AA. In the $z<1.65$ DLA catalog of \citet{rao2017}, there are no absorbers with log~$N$(\HI) $\ge$ 19.75 and W$_r$(2796) $\le$ 0.12\,\AA\ (as the absorber detected at $z=0.5305$ in this study). We note that even though the spectroscopic targets in the study of \citet{rao2017} were picked based on the presence of strong \MgII\ absorption, the survey itself is not biased against detecting DLAs with no or weak \MgII\ absorption along these sightlines. Similarly, there are no absorbers with the $N$(\HI) and W$_r$(2796) constraints as the $z=0.5305$ reported here in the high-$z$ ($2<z<6$) sample studied by \citet{Matejek2013}. 

Using the upper limit on the \MgII\ rest equivalent width for this system from the DESI spectra (i.e., 0.12\,\AA) and assuming the linear part of the curve of growth, we get an upper limit on the column density of Mg~{\sc ii} of $\sim2.8\times 10^{12}$ cm$^{-2}$. If we use the $N$(\HI) inferred for $T_S = 100$ K and $f_c = 1$ in Table~\ref{tab:21cm}, we get an upper limit on the metallicity of $\sim1.5\times10^{-3}$ Z$_\odot$. Given the fact that the covering factor can be lower than 1, the measured spin temperatures in the high-$z$ absorbers are usually higher than 100 K \citep[in particular if we consider the low metallicity, as suggested by][]{Kanekar09ts}, the actual metallicity could be even lower than the above value.
That will make this system an extremely metal poor DLA. Such systems are very rare but useful for probing the early enrichment history of DLA host galaxies \citep[see][and references therein]{Welsh2024}. In particular, no such metal-poor DLA is known at $z\sim0.6$ \citep[see Table 4 of][for example]{Rahmani2016}.  All this makes this an interesting system for future detailed investigations.

Alternatively, the lack of \MgII\ absorption could be related to the optical and radio lines of sight being different. This is possible if the offset ($\sim 1$ arcsec or $\sim 6$ kpc) seen between the peak radio emission in the L-band image and the optical position is indeed true.
In this scenario, we require a steep gradient in the \MgII\ equivalent width (and/or \HI\ column density) of the gas producing \HI\ 21-cm absorption. 
Such variations in the absorbing gas have been reported in studies of gravitationally lensed quasars.
\citet{Ellison2004} have reported that low-ionization systems (primarily \MgII\ absorbers) show strong equivalent width variations (often $>80\%$) over kpc scales using the HST spectra of gravitationally lensed quasar APM~08279+5255. 
\citet{Okoshi2021} found that inhomogeneous cold absorbers that give rise to strong \HI/\MgII\ absorption reside on a scale of about 6-12 kpc between the four lines of sight of the gravitationally lensed quasar H1413+1143. It was found that the \HI\ column density and \MgII\ equivalent width in this case typically change by factors of 2-40 and $\sim$6, respectively, over transverse separations of 6-12 kpc \citep[see also][]{Monier2009,Augustin2021,Bordoloi2022,Dutta2024}. On the other hand, \citet{Rubin2018lens} have found much larger coherence scales, with \MgII\ equivalent width varying by $\le$40\% over the transverse scale of 8-22 kpc along the lines of sight toward the quadruply lensed quasar J014710+463040. 
Furthermore, using \HI\ 21-cm absorption towards an extended background radio source and \caii\ and \nai\ absorption towards the optical quasar, \citet{Dutta16} have revealed sub-kpc- and kpc-scale variations in the neutral gas in the outskirts of a low redshift galaxy.
Thus, a possible presence of large column density variation in \HI\ (or equivalent width variation of \MgII) over kpc scales in the present case is probable. Low frequency arcsec and sub-arcsec radio images are needed to confirm this possibility. Alternatively, detection of time-variability of \HI\ 21-cm absorption \citep[as in][]{Srianand2022} will also favor such a picture.

\subsection{Probability of occurrence of multiple \HI\ 21-cm absorbers}

To date, the closest \HI\ 21-cm absorption systems along a single sightline are \zhi\ = $0.09123$ and $0.22125$, toward OI 363 (B2 0738+313) reported by \citet{Kanekar03}. These absorbers were pre-selected for \HI\ 21-cm searches based on the presence of DLA systems \citep{Rao1998}. In contrast, the velocity separation between the two absorbers in our case is significantly smaller (i.e., $\Delta z = 0.0348$) than that observed toward OI 363.

Blind searches for \HI\ 21-cm absorption reported to date usually lead to a very small number of detections \citep{Allison2020, Sadler2020, Gupta2021, Hu2025, Yoon25}. A recent blind deep MeerKAT search conducted as part of the Looking At the Distant Universe with the MeerKAT Array \citep[LADUMA,][]{Blyth2016} survey found \HI\ 21-cm absorption detection rate of $\sim0.3\%$ at $z<0.62$, including detections toward some of the lowest flux density sources probed to date (M. Glowacki et al. 2026, in preparation). Therefore, estimating the precise probability of occurrence of two \HI\ 21-cm absorbers as we have reported here from an unbiased survey is difficult at this stage. Based on the number of intervening \HI\ 21-cm absorbers per unit redshift provided by \citet{Sadler2020}, we estimate the random probability of occurrence of a second \HI\ 21-cm absorber at $\Delta z\le0.035$ to be 0.007$^{+0.005}_{-0.003}$ at $z\sim0.6$.

Next, we try to estimate the probability of occurrence of two DLAs with $\Delta z\le 0.035$ along a single line of sight. For this, we use the DLA catalog of \citet{rao2017}. In this study, the Hubble Space Telescope (HST) spectra were obtained to search for DLAs at $z<1.65$ towards UV bright quasars that are known to have \MgII\ absorbers (W$_r$(2796) $\ge$ 0.3\AA) along their line of sight. The number of sightlines showing two DLAs within a redshift interval of 0.035 is 2 out of 239 sightlines probed. Therefore, the probability of detecting two DLAs along the same line of sight within the redshift interval $\Delta z\sim 0.035$ is $\sim$0.8\%. Since not all DLAs will produce \HI\ 21-cm absorption \citep[e.g., $\approx$60\% of $z<1$ DLAs show \HI\ 21-cm absorption;][]{Dutta2017}, this should be considered as the upper limit on the coincidence probability. 

\begin{figure*}
\begin{center}
\includegraphics[width=3.5in, trim=0cm 0cm 0cm 0cm, clip]{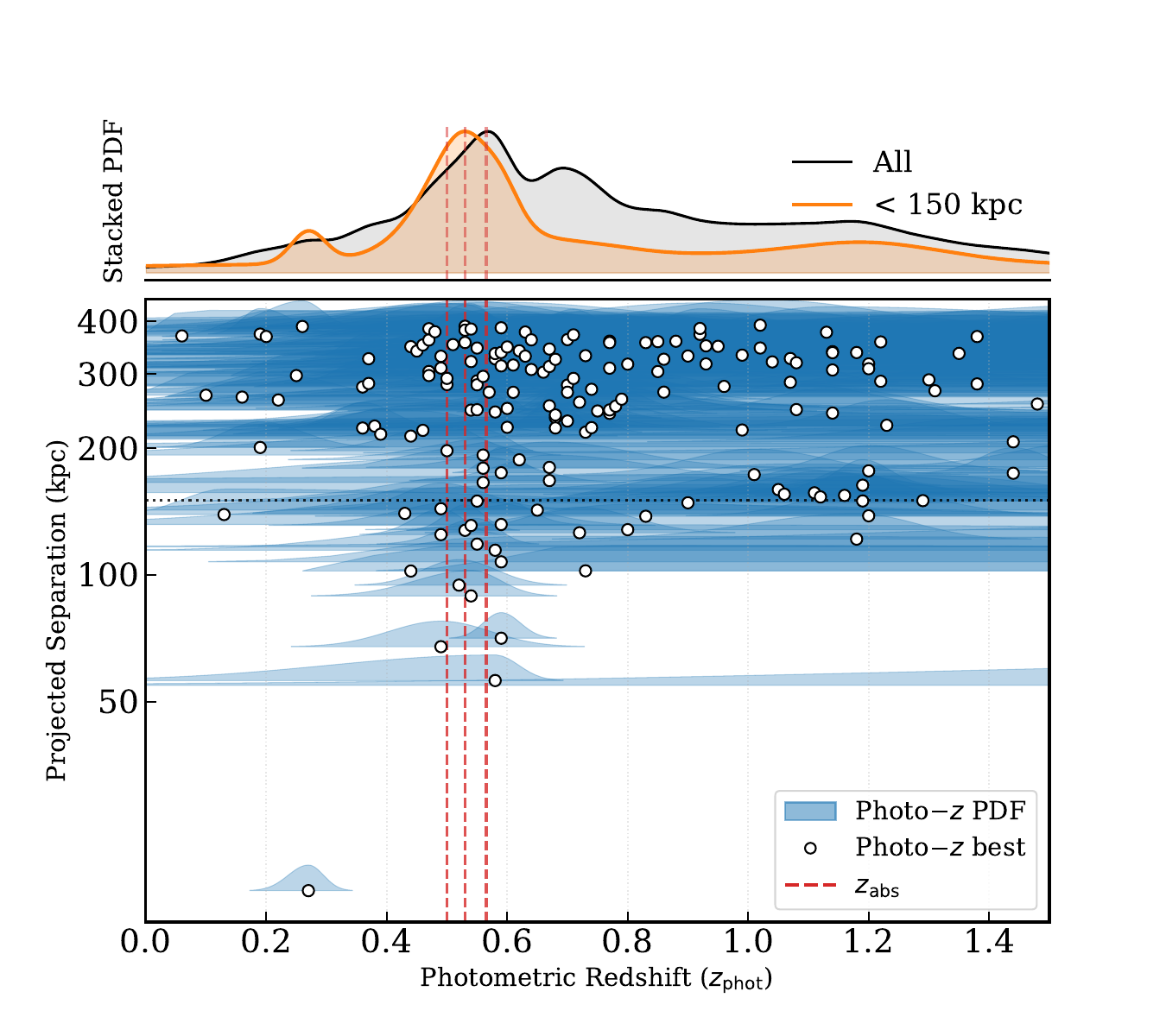}
\includegraphics[width=3.5in, trim=0cm 0cm 0cm 0cm, clip]{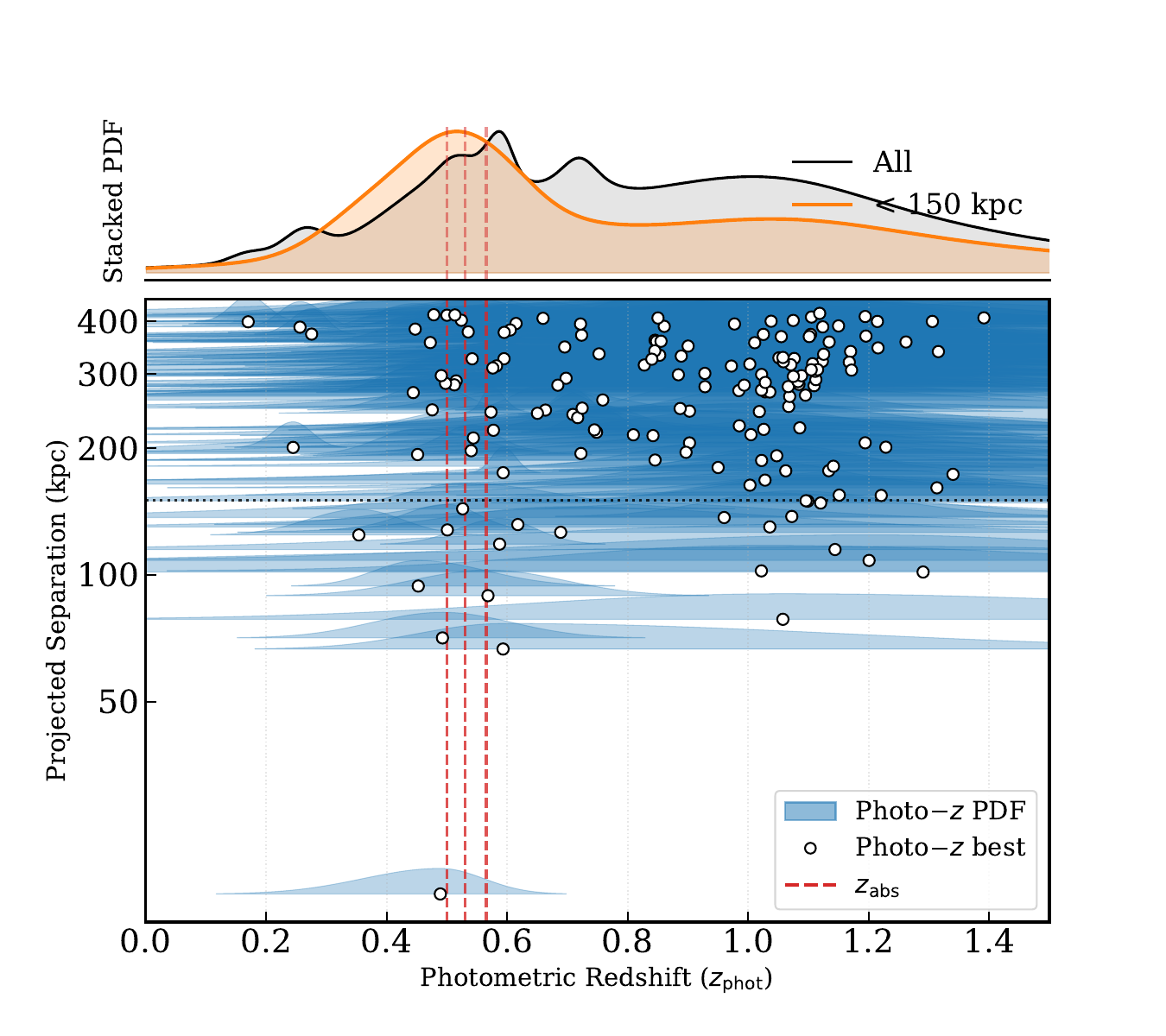}
\caption{Photometric redshift analysis of galaxies within a $1^\prime$ field of view centered on the quasar J1201+0028. The panels compare data from HSC (left) and DESI LS10 (right). Lower panels: Projected physical separation as a function of best-fit photometric redshift, with horizontal bars indicating the 68\% confidence intervals. Upper panels: The corresponding stacked photometric redshift PDF. Vertical dashed lines mark the redshifts of all the absorption systems (\zabs\ $=0.50046$, 0.5305, and 0.56544) detected along the sightline.}
   \label{photo-z-hsc}
\end{center}
\end{figure*}

Next, we estimate the coincidence probability using \MgII\ absorbers. We constructed a catalog of \MgII\ absorbers at $0.45 \le z \le 0.65$ from the SDSS \MgII\ catalog of \citet{Zhu2013}, which is typically $>80\%$ complete for W$_r(2796)\ge1$\AA\ (see their Figure 7). To prevent the kinematic substructure of a single absorption system from being misidentified as individual absorption systems, we merged components along each sightline separated by $\Delta v \le 500$~km~s$^{-1}$ using a one-dimensional Friends-of-Friends (FoF) algorithm and adopted their mean redshift as the system centroid. We then identified sightlines hosting multiple \MgII\ systems within $\Delta z \le 0.1$. This yields 58 systems along 468 unique sightlines, corresponding to an incidence of $\sim$12.4\%. For the full \citet{Zhu2013} sample, the corresponding incidence rate is $\sim$21.9\%.
Assuming a $\sim$10\% detection probability of \HI\ 21-cm absorption in \MgII\ systems \citep{Dutta17mg2}, we estimate a coincidence probability of order $\sim$0.1\% for detecting two \HI\ 21-cm absorbers within the redshift range of our interest along a single sightline. Considering the full redshift range probed by the \citet{Zhu2013} sample, this probability rises to $\sim$0.2\%.

Therefore, a chance coincidence of two \HI\ 21-cm absorbers with one of them showing no \mgii\ absorption could be related to this line of sight passing through some overdense region of galaxies with varied metal enrichment. We explore this further below.
\subsection{Galaxy overdensity as possible origin of the coincidence absorbers}

The available Hyper Suprime-Cam (HSC) image suggests two possible galaxies within a projected separation of $b \lesssim 20$~kpc from the quasar sightline if they are at the redshift of the absorbers. The galaxy located at $b \approx 17$~kpc has a DESI photometric redshift of $z_{\rm phot} = 0.49_{-0.13}^{+0.07}$ (68\% confidence), which is consistent with both the \HI\ 21-cm absorption systems. The second galaxy, located in front of the quasar, lacks spectroscopic or photometric redshift constraints due to severe contamination from the bright quasar PSF. 

To investigate further, 
we retrieved photometric redshifts ($z_{\rm phot}$) from two complementary catalogs: the HSC survey PDR3 \citep{Aihara22}
($5\sigma$ point-source depth, $g = 26.5^{+0.2}_{-0.2}$~mag)
and the DESI Legacy Imaging Surveys \citep[DESI LS10;][]{Dey2019}
$5\sigma$ point-source depth, $g \simeq 24.7$~mag).
We performed a radial search within a radius of $\sim 1.3^\prime$ around the quasar line-of-sight, which corresponds to a projected physical separation of approximately 500~kpc at the mean redshift of the absorbers ($z \approx 0.53$).
Figure~\ref{photo-z-hsc} shows the redshift distributions of the galaxies, with the HSC sample (211 galaxies) shown in the left panel and the DESI LS10 sample (165 galaxies) shown in the right panel. 
The main panels in Figure~\ref{photo-z-hsc} display the individual probability density function (PDF) of redshift as ``ridgeline'' plots.
%
The top panels present the cumulative probability distributions, created by stacking the individual PDFs. We show both the total population (black) and the subset of galaxies located within an impact parameter of $b < 150$~kpc (orange: inner region), to highlight the contribution of the innermost sources.

The vertical dashed lines in Figure~\ref{photo-z-hsc} indicate the redshifts of the three distinct absorption systems of interest here ($z_{\rm abs} = 0.50046, 0.5305, 0.56544$). As seen in the top panels, the stacked photometric PDF from both HSC and DESI LS10 reveals a broad but significant galaxy overdensity coincident with the redshift range of the $z\sim0.5$ absorbers.
Notably, when isolating the closer galaxy population (projected separation $< 150$~kpc, orange curve), the probability distribution in both datasets narrows and peaks more prominently near the absorber redshifts.

For the \HI\ 21-cm absorber at \zabs\ $= 0.5305$, applying a hard selection criterion requiring a $1\sigma$ overlap between the galaxy photometric redshift and the absorber redshift yields 79 (45) candidate galaxies within a projected radius of 500 kpc for the HSC and DESI catalogs (DESI values are given in parentheses).
Of these, 16 (12) lie within a projected separation of 150 kpc.
Next, we integrate the individual redshift PDFs within $\Delta z = \pm 0.04$ of the absorber redshift to estimate the statistically expected number of associated galaxies. This method yields an effective count of $\sim 19.2$ (13.6) galaxies across the full field and $\sim 4.7$ (2.7) within the inner 150 kpc region. 
We estimate the expected galaxy number density by integrating a Schechter luminosity function in the SDSS $g-$band from \citet{Blanton2003} down to the limiting absolute magnitude appropriate for the HSC and DESI data.
Comparing the observed counts to these expectations, we derive overdensity values of $\delta \sim 2.3$ and $\sim 3.4$ within a 500 kpc radius for HSC and DESI, respectively.

We find a comparable environment for the \zabs\ $= 0.5653$ absorber.
In this case, 79 (50) candidate galaxies are identified within a 500 kpc radius using the hard selection, with 15 (10) located inside 150 kpc.
Applying the same probabilistic PDF integration method as above yields effective counts of $\sim 20.8$ (14.7) galaxies across the full field and $\sim 4.4$ (2.5) within the inner region.
The corresponding overdensity values are $\delta \sim 2.3$ and $\sim 3.6$ for HSC and DESI, respectively.
These consistently elevated overdensities indicate that both \HI\ 21-cm absorbers reside in significantly enhanced galaxy environments, consistent with association with group-scale structures.

The advent of integral field spectroscopic (IFS) observations have revealed the association of a significant fraction of absorbers detected in optical spectra of quasars with overdense structures such as galaxy groups \citep[e.g.,][]{Hamanowicz2020,Dutta2021,Qu2023,Galbiati2024}. While the galaxy environment around intervening \HI\ 21-cm absorbers has not been extensively studied, there have been a few studies that have reported \HI\ 21-cm absorption associated with galaxy groups \citep[e.g.,][]{Dutta2016,Peroux2019,Weng2022,Maina2022}. In the present case, the substantial spread in redshift ($\Delta z \approx 0.04$) of the galaxy overdensity associated with the absorbers at $z\sim0.5$ likely exceeds the intrinsic velocity dispersion of a single virialized cluster. Instead, this spread implies the presence of a large-scale structure extended along the line of sight, or the superposition of multiple clustering nodes within a broader cosmic web filament. While there have been a few recent detections of large-scale structures or cosmic web filaments in emission \citep[e.g.,][]{Umehata2019,Bacon2021,Tornotti2025,Banerjee2025,Arabsalmani2025}, the present study indicates that blind absorption line surveys could present a complementary technique of identifying such structures.

While the discussion presented here suggests that a possible galaxy overdensity could be responsible for the coincident absorption, definitive progress in linking the absorber properties to their underlying galaxy environment will require deep IFS observations. In addition, wide-area spectroscopic coverage is essential to trace the surrounding large-scale structure and to robustly assess the role of environment on the observed absorption.

\begin{acknowledgments}
The MeerKAT telescope is operated by the South African Radio Astronomy Observatory, which is a facility of the National Research Foundation, an agency of the Department of Science and Innovation. We acknowledge use of the IDIA data intensive research cloud for data processing. The IDIA is a South African university partnership involving the University of Cape Town, the University of Pretoria and the University of the Western Cape.
The MeerKAT data were processed using the {\it ilifu} cloud computing and IUCAA computational facilities.
AM acknowledges support from the South African Radio Astronomy Observatory (SARAO) towards this research and the Inter-University Centre for Astronomy and Astrophysics (IUCAA), India
for providing hospitality and travel grants while carrying out this work. The authors thank the anonymous referee for a constructive and helpful report.
The Common Astronomy Software Applications (CASA) package is developed by an international consortium of scientists based at the National Radio Astronomical Observatory (NRAO), the European Southern Observatory (ESO), the National Astronomical Observatory of Japan (NAOJ), the Academia Sinica Institute of Astronomy and Astrophysics (ASIAA), the CSIRO division for Astronomy and Space Science (CASS), and the Netherlands Institute for Radio Astronomy (ASTRON) under the guidance of NRAO.
\facilities{MeerKAT}

This work makes use of the following softwares and packages: \textsc{casa} \citep{CASA2022}, \textsc{aoflagger} \citep{Offringa2012}, \textsc{numpy} \citep{numpy2020}, \textsc{matplotlib} \citep{matplotlib2007}, \textsc{scipy} \citep{scipy2020}, \textsc{astropy} \citep{astropy:2013,astropy:2018,astropy:2022}, \textsc{WSClean} \citep{Offringa2014}.
\end{acknowledgments}




\bibliography{newbib}{}
\bibliographystyle{aasjournal}

\appendix

\section{Optical spectrum towards J1201$+$0028}
\label{sec:apendix}
\begin{figure}[htbp]
    \centering
    \includegraphics[width=0.45\linewidth]{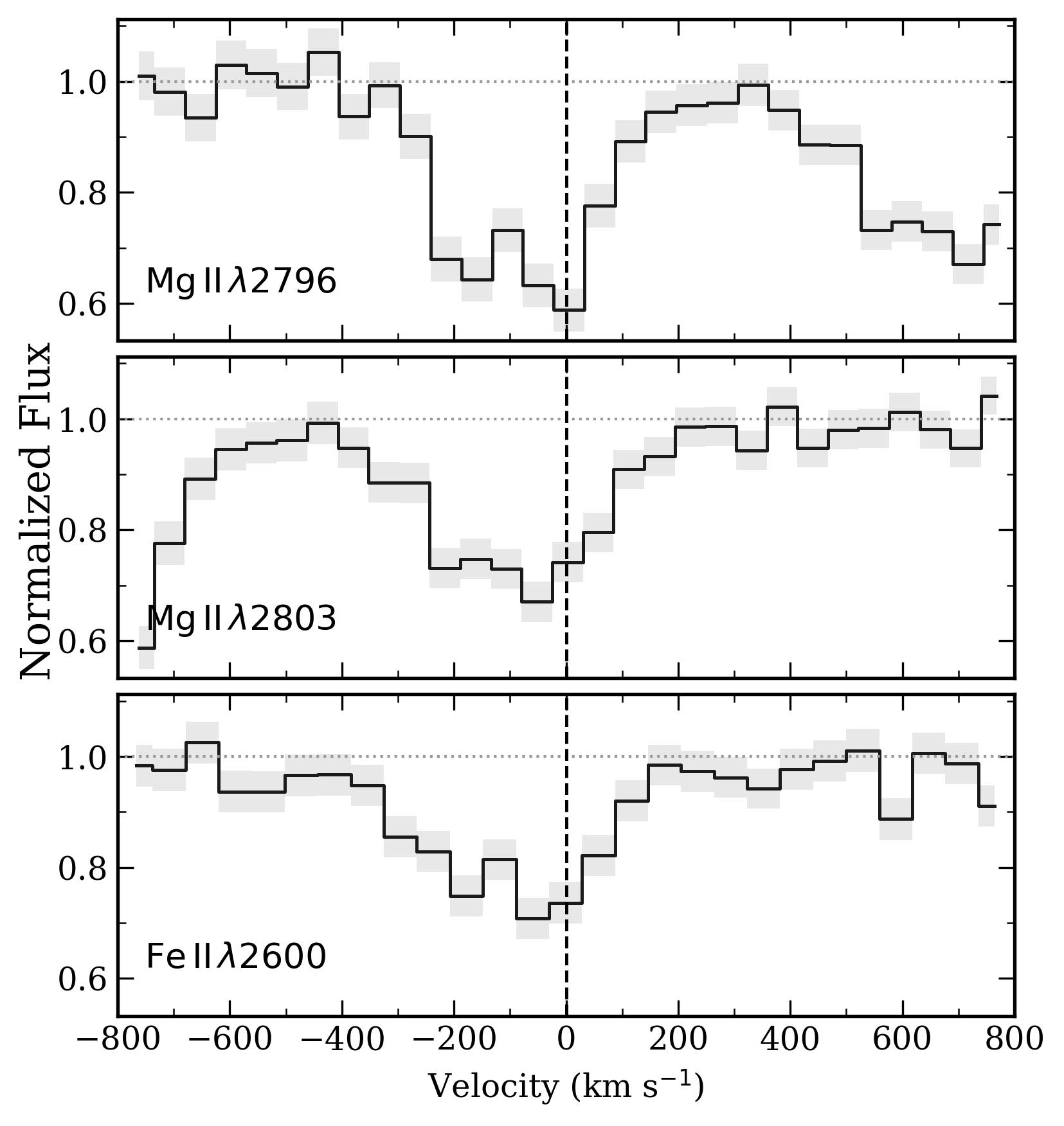}
    \caption{Normalized flux vs velocity of \mgii\, $\lambda2796$, \mgii\, $\lambda2803$, and \feii\, $\lambda2600$ at $z_{\rm abs} = 0.56544$. The gray shaded regions denote the $1\sigma$ flux uncertainties. The vertical dashed line marks zero velocity relative to the absorber redshift.}
    \label{fig:optical}
\end{figure}

\section{Robustness of the \HI\ 21-cm Absorption Detections}
\label{apexb}
We performed a series of tests to assess the robustness of the two \HI\ 21-cm absorption detections against RFI. We inspected the absorption features in the two
parallel-hand polarization products (XX and YY), and performed
spatially offset spectral extractions around the target position.
These tests are described below.
\subsection{Consistency of the Absorption Features}

\begin{figure*}
\begin{center}
\includegraphics[width=3.1in, trim=0cm 0cm 0cm 0cm, clip]{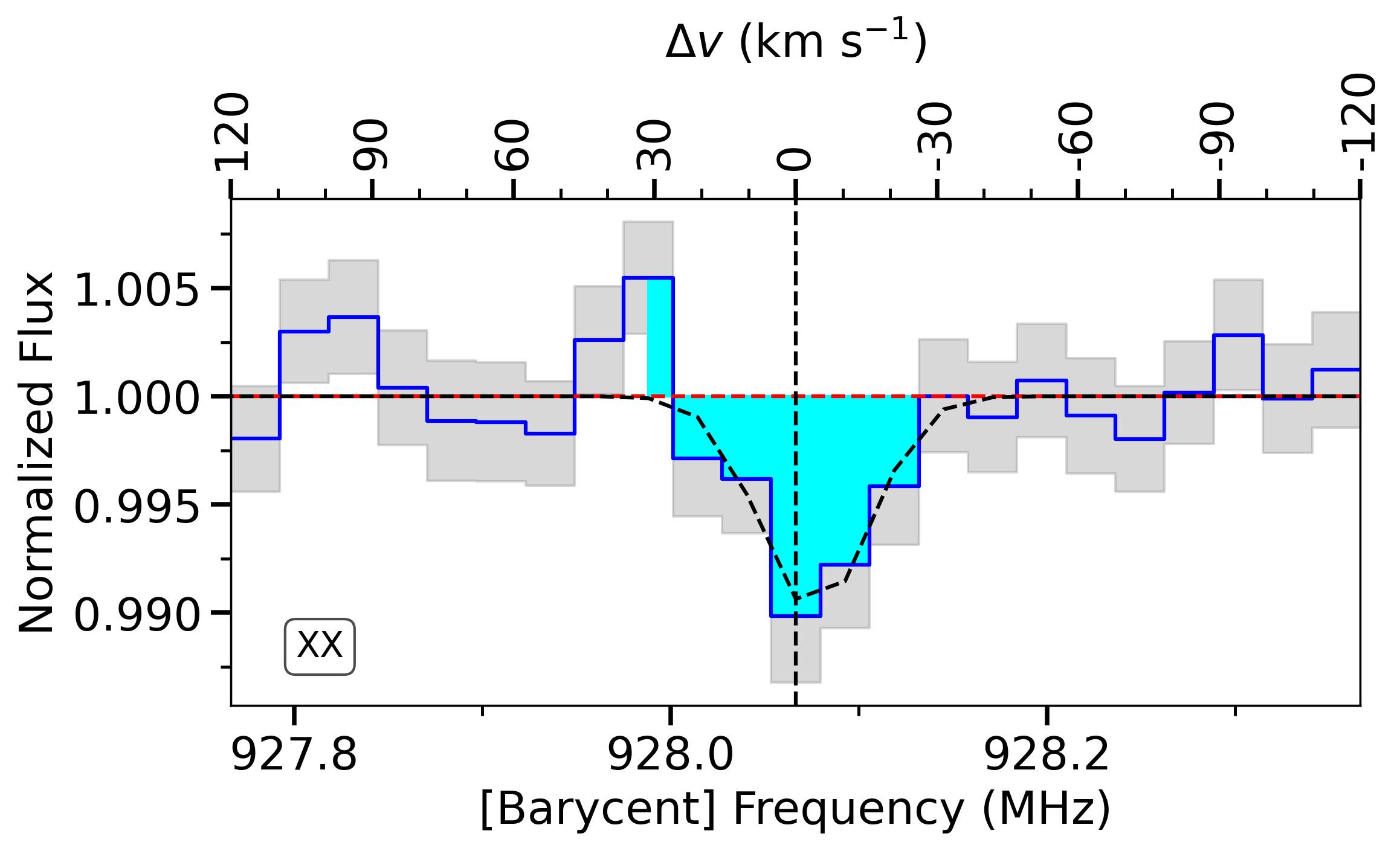}
 \includegraphics[width=3.1in, trim=0cm 0cm 0cm 0cm, clip]{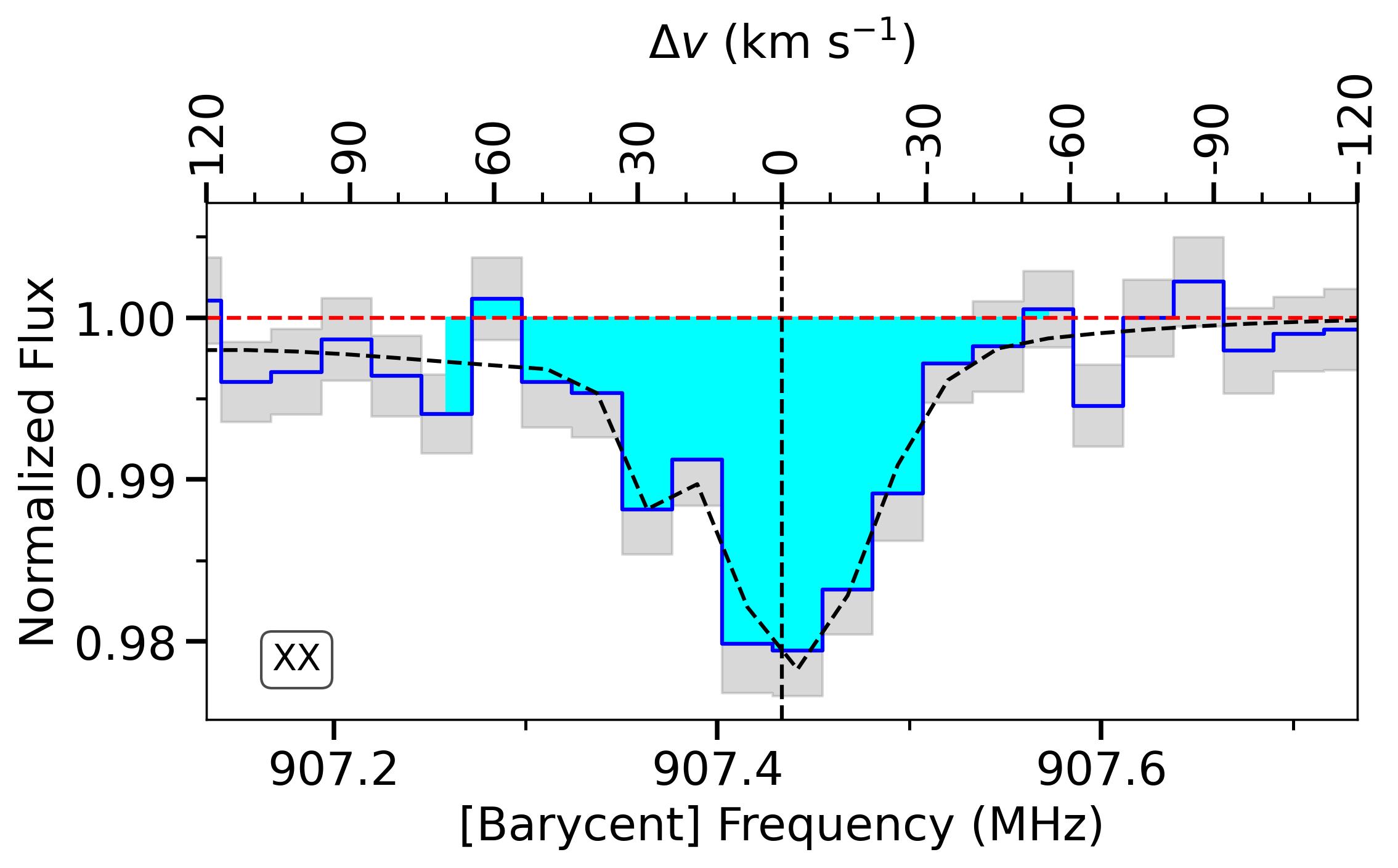}
 \includegraphics[width=3.1in, trim=0cm 0cm 0cm 0cm, clip]{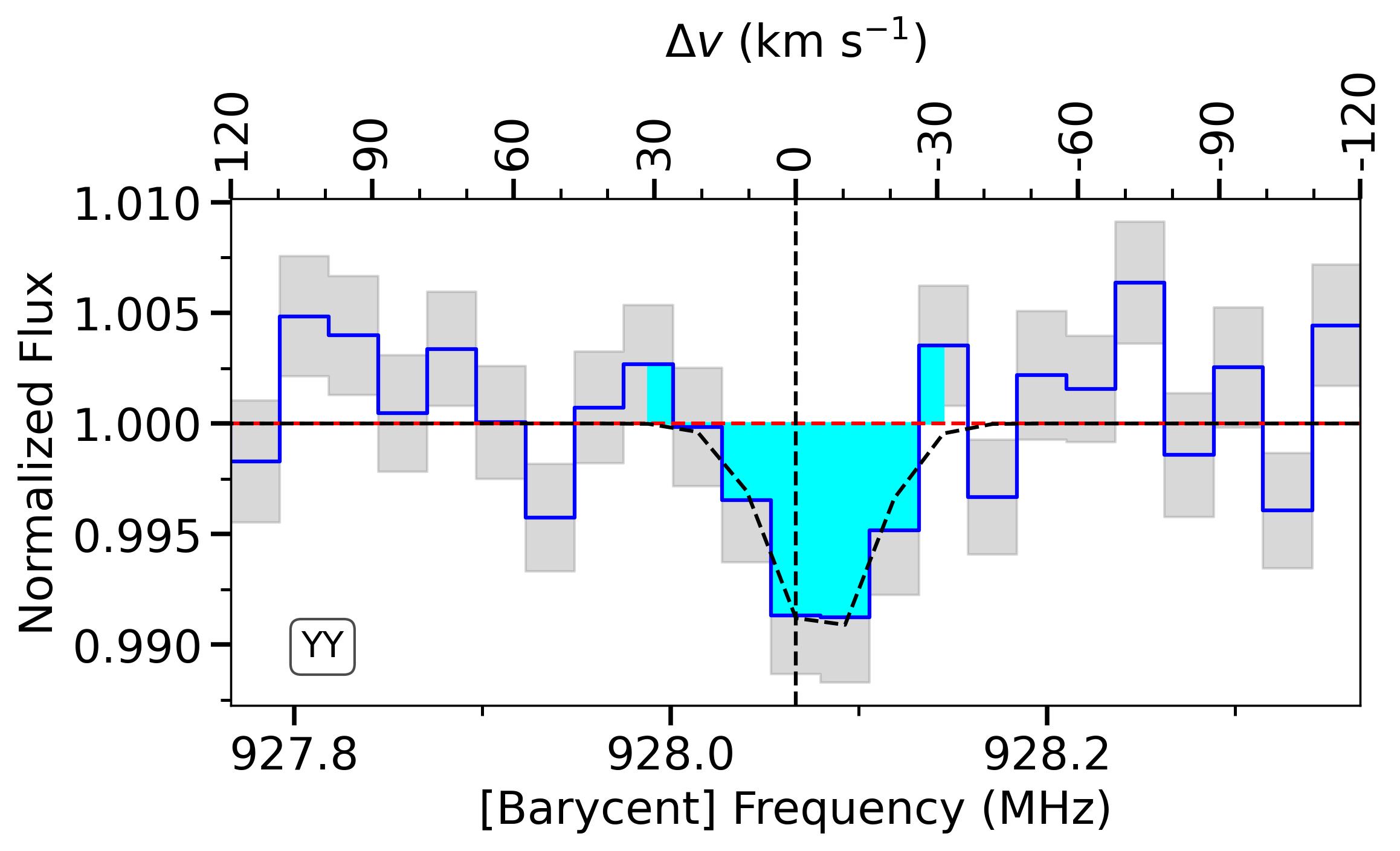}
 \includegraphics[width=3.1in, trim=0cm 0cm 0cm 0cm, clip]{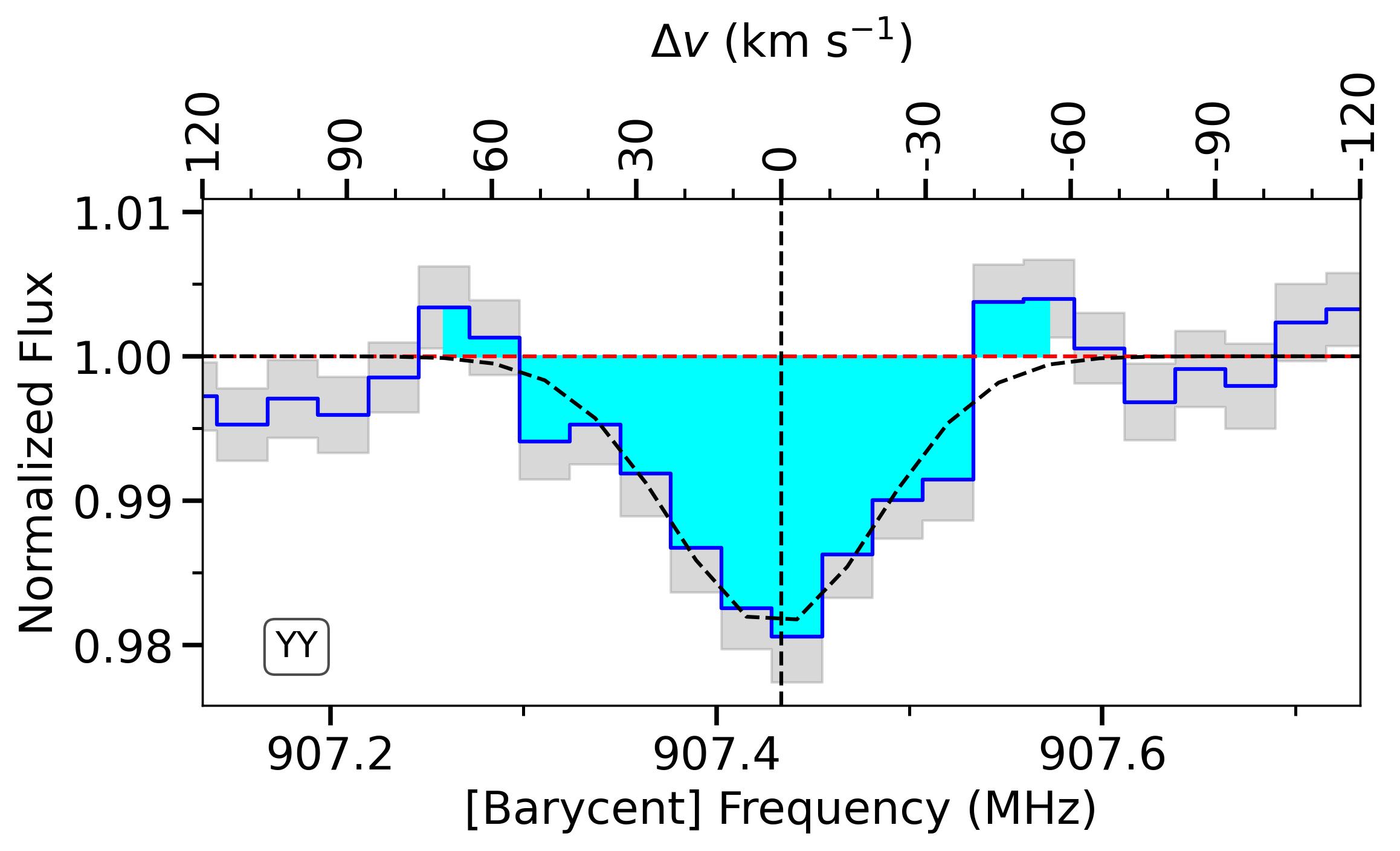}
 \caption{Consistency of the \HI\ 21-cm absorption features (shown in native resolution) in the
    two parallel-hand polarization products. The left and right
    columns show the absorbers at \zhi\ = 0.5305 and
    \zhi\ =0.5653, respectively. The upper and lower panels
    show the continuum-normalized XX and YY spectra, respectively.
    The gray shaded regions indicate the $1\sigma$ uncertainties.
    The cyan shaded regions indicate the velocity intervals used to
    calculate the integrated optical depth. The black dashed curves
    show the corresponding best-fitting absorption profiles, while
    the red horizontal dashed lines indicate the normalized
    continuum level. The upper axes show the velocity offset
    relative to the peak \HI\ 21-cm optical depth. The absorption
    features occur at consistent velocities in both polarization
    products, demonstrating their presence independently in XX and
    YY.
    }
\label{fig:xxyy}
\end{center}
\end{figure*}

We examined the absorption features in the two
parallel-hand polarization products, XX and YY, to test whether the
detections could be associated with any RFI. Figure~\ref{fig:xxyy} shows
the continuum-normalized XX and YY spectra around the two \HI\
21-cm absorption systems. The absorption features are clearly detected in both polarization products and occur at consistent velocities. For the absorber at $z_{\rm HI}=0.5305$, the integrated optical depths
measured independently from the XX and YY spectra are
$\int\tau_{\rm XX}dv = 0.28\pm0.08$ and
$\int\tau_{\rm YY}dv =  0.24\pm0.08$~\kms,
respectively. For the absorber at $z_{\rm HI}=0.5653$, the
corresponding values are
$\int\tau_{\rm XX}dv = 1.24\pm 0.08$ and
$\int\tau_{\rm YY}dv = 1.17\pm0.08 $~\kms.
These measurements are consistent and are above the $ 3\sigma$ level.
\subsection{Spatial Extraction Test}
\begin{figure*}
\begin{center}
\includegraphics[width=7in, trim=0cm 0cm 0cm 0cm, clip]{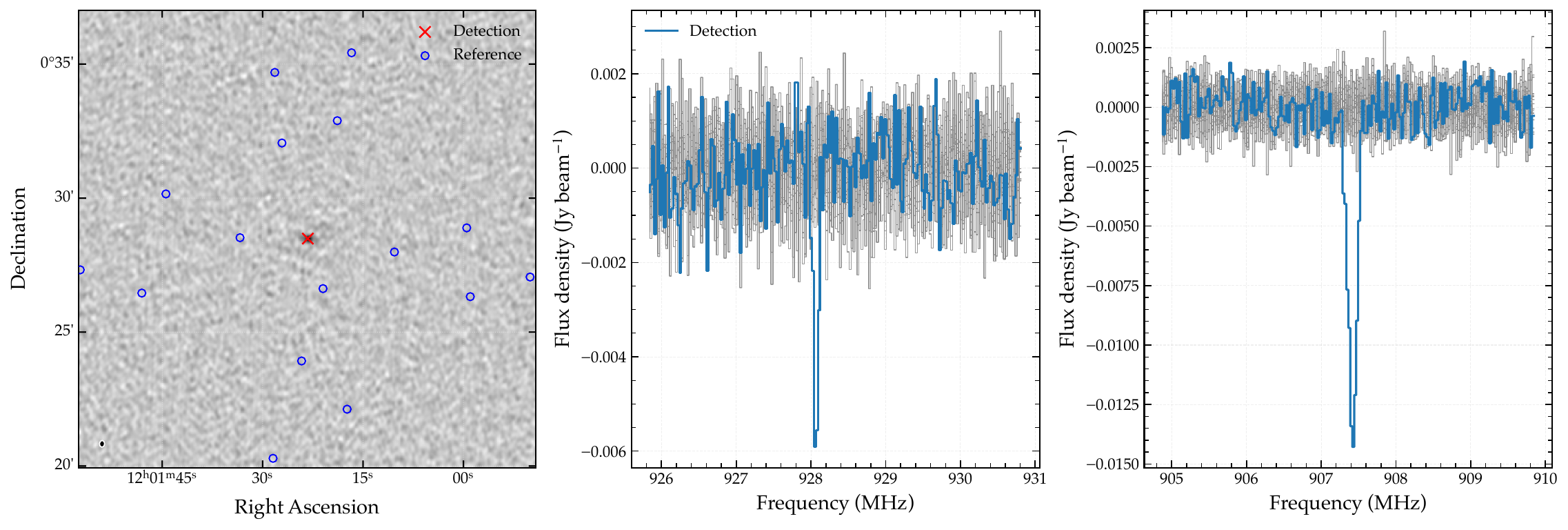}
\caption{Spatial extraction test for the \HI\ 21-cm absorption
    features towards J1201+0028. The left panel shows the channel image (beam is shown in lower left corner) with the position of the \HI\ 21-cm detection
    marked by a red cross and the nearby reference positions marked
    by blue circles. The middle and right panels show the spectra
    extracted at the detection position (blue) and at the reference
    positions (gray) around the $z_{\rm HI}=0.5305$ and
    $z_{\rm HI}=0.5653$ absorbers, respectively.}
\label{fig:spatial_extraction}
\end{center}
\end{figure*}

As an additional test of the origin of the absorption features, we
performed spectral extractions at the position of the absorption and at
multiple random nearby reference positions. The detection position and the
reference positions are shown in the left panel of
Figure~\ref{fig:spatial_extraction}. The extracted spectra are shown in the middle and right panels of
Figure~\ref{fig:spatial_extraction}, covering the frequency ranges
around the absorbers at $z_{\rm HI}=0.5305$ and $z_{\rm HI}=0.5653$,
respectively. The absorption features are present in the spectrum
extracted at the target position, whereas no absorption
feature is seen in the spectra extracted at the reference
positions. 



\end{document}